# Multi-periodic pulsations of a stripped red giant star in an eclipsing binary

Pierre F. L. Maxted[1], Aldo M. Serenelli[2], Andrea Miglio[3], Thomas R. Marsh[4], Ulrich Heber[5], Vikram S. Dhillon[6], Stuart Littlefair[6], Chris Copperwheat[7], Barry Smalley[1], Elmé Breedt[4], Veronika Schaffenroth[5,8]

[1] Astrophysics Group, Keele University, Keele, Staffordshire, ST5 5BG, UK.

[2] Instituto de Ciencias del Espacio (CSIC-IEEC), Facultad de Ciencias, Campus UAB, 08193, Bellaterra, Spain.

[3] School of Physics and Astronomy, University of Birmingham, Edgbaston, Birmingham, B15 2TT, UK.

[4] Department of Physics, University of Warwick, Coventry, CV4 7AL, UK.

[5] Dr. Karl Remeis-Observatory & ECAP, Astronomical Institute, Friedrich-Alexander University Erlangen-Nuremberg, Sternwartstr. 7, 96049, Bamberg, Germany.

[6] Department of Physics and Astronomy, University of Sheffield, Sheffield, S3 7RH, UK.

[7] Astrophysics Research Institute, Liverpool John Moores University, Twelve Quays House, Egerton Wharf, Birkenhead, Wirral, CH41 1LD, UK.

[8] Institute for Astro- and Particle Physics, University of Innsbruck, Technikerstrasse. 25/8, 6020 Innsbruck, Austria.

Low mass white dwarfs are the remnants of disrupted red giant stars in binary millisecond pulsars[1] and other exotic binary star systems[2–4]. Some low mass white dwarfs cool rapidly, while others stay bright for millions of years due to stable **fusion in thick surface hydrogen layers[5]. This dichotomy is not well understood so their potential use as independent clocks to test the spin-down ages of pulsars[6,7] or as probes of the extreme environments in which low mass white dwarfs form[8–10] cannot be fully exploited. Here we present precise mass and radius measurements for the precursor to a low mass white dwarf. We find that only models in which this star has a thick hydrogen envelope can match the strong constraints provided by our new observations. Very cool low mass white dwarfs must therefore have lost their thick hydrogen envelopes by irradiation from pulsar companions[11,12] or by episodes of unstable hydrogen fusion (shell flashes). We also find that this low mass white dwarf precursor is a new type of pulsating star. The observed pulsation frequencies are sensitive to internal processes that determine whether this star will undergo shell flashes.**

Most white dwarfs have no internal energy sources and the mass of hydrogen on their surfaces is small ($\lesssim 10^{-4}$ solar masses) so their "cooling age" can be accurately estimated from their current luminosity. The predicted mass range for which shell flashes at the base of the hydrogen envelope occur in low mass white dwarfs depends on the assumed composition and whether diffusion is included in the model[13–15], so it is generally possible to find a scenario for a particular binary millisecond pulsar in which the white dwarf cooling age is consistent with the pulsar spin down age. However, recent observations have shown that the white dwarf companion to the pulsar PSR J0751+1807 is much too cool to have a thick hydrogen envelope, but the white dwarf is not massive enough to have undergone shell flashes and there is no evidence for strong irradiation by the pulsar[16]. The standard assumptions used to derive pulsar spin-down ages have also been called into question[17,18]. A better understanding of the initial hydrogen layer mass and its evolution in low mass white dwarfs is needed so that we can make an independent test of the spin-down ages for pulsars.

Prior to becoming low mass white dwarfs, stripped red giants evolve at nearly constant luminosity towards higher effective temperatures. 1SWASP J024743.37-251549.2 (J0247-25 hereafter) was recently discovered to be a binary system in which a star in this rarely observed evolutionary phase (J0247-25B) is totally eclipsed by its companion star (J0247-25A)[19]. We have obtained new spectroscopic and photometric observations (Fig. 1) and used these to derive precise astrophysical parameters for both stars (Table 1).

The mass, radius and luminosity of J0247-25A are well matched by models of stars with a metal abundance $Z = 0.004 - 0.019$, but not for models outside this range (Fig. 2). We have calculated models for the formation of J0247-25B by mass transfer onto the companion star (Fig. 2). We assumed that the mass loss rate is slower than the thermal timescale of the star, so mass transfer ceases when the equilibrium radius of the star is smaller than the Roche lobe. This is the assumption usually made in the absence of a detailed understanding of the mass loss process[5,10,15,20]. Diffusion of elements by gravitational settling, chemical diffusion and thermal diffusion are included in our models. We also calculated models without diffusion to investigate the effects of processes such as rotation that may counteract diffusion. The mass of J0247-25B is near the lower limit for the occurrence of shell flashes – these will occur if its metal abundance is high enough ($Z \approx 0.01$) and diffusion can increase the hydrogen abundance in the regions where shell flashes can be initiated.

There is a direct relationship between orbital period, mass and composition that arises from the assumption that the mass-losing star is in equilibrium when mass transfer ceases[20]. We find a good match to the observed orbital period of J0247-25 for models with a similar range of metal abundance as J0247-25A, but not for models outside this composition range (Supplementary Table 5). This result is not strongly affected by the assumptions made about the evolution of the binary system during the mass transfer episode.

All the models that provide a good match to the observed properties of J0247-25B have thick hydrogen envelopes ($\approx 0.005$ solar masses). We did produce models for J0247-25B with thin hydrogen envelopes but found that for any reasonable estimate of the composition these models never have properties like J0247-25B. Given the strong observational constraints on the mass, radius, luminosity, age, orbital period and composition for J0247-25 and the pulsation properties predicted by these models discussed below, we conclude that only models in which low mass white dwarfs are born with thick hydrogen envelopes can match the observed properties of J0247-25B.

J0247-25B pulsates with at least 3 different frequencies around 2500μHz (Supplementary Fig. 2, Supplementary Table 3). We have assessed the potential of this new class of pulsating star for asteroseismology, i.e., to study the interior structure of stripped red giants from an analysis of their pulsation frequencies. We have used the adiabatic approximation to calculate potential oscillation frequencies for our models of J0247-25B (Fig. 3). These calculations show that the observed frequencies are a mixture of radial modes and non-radial modes. The identification of which frequencies are radial modes and the angular degree, ℓ, for the non-radial modes is required before the observed frequencies can be used for asteroseismology. Standard techniques exist to identify modes, although these identifications can be model dependent. The identification of modes from the analysis of the eclipses in an eclipsing binary like J0247-25 provides a rare opportunity to test standard mode identification techniques[21,22].

 The non-radial modes in J0247-25B behave like gravity modes near the core and like pressure modes in the outer layers of the star. These mixed modes can be used to study the entire structure of the star, e.g., to measure its internal rotation profile[23,24]. Information on the interior structure of J0247-25B is also contained in the radial modes (Fig. 3).  The pulsation periods are comparable to the thermal relaxation timescale in the second partial ionisation zone of helium so the change in opacity in this region acting like a heat engine may be driving these oscillations (κ-mechanism). This suggests that other low mass white dwarf precursors[25] with effective temperatures similar to J0247-25B may also show pulsations. Thus, the discovery of pulsations in J0247-25B opens up the possibility for a much-improved understanding of the structure and formation of low mass white dwarfs and their environments through the application of asteroseismology to this new class of pulsating variable stars.

**Supplementary Information** accompanies the paper on www.nature.com/nature.

**Acknowledgements** We thank the ESO staff who obtained our UVES data for taking care to schedule the observations at the correct orbital phases. We thank Aleksey Cherman and Don Kurtz for comments on a draft version of the paper. Based on observations collected at the European Southern Observatory, Chile (Program ID: 086.D-0194). A.M.S. is partially supported by the Re-integration Grant PIRG-GA-2009-247732 (FP7-People), and the MICINN grant AYA2011-24704. V.S. acknowledges funding by the Deutsches Zentrum für Luft- und Raumfahrt (grant 50 OR 1110) and by the Erika-Giehrl-Stiftung. T.R.M. acknowledges funding from the United Kingdom's Science and Technology Facilities Council (ST/I001719/1)

**Author Contributions** P.F.L.M. analysed the lightcurves and spectroscopy and wrote the paper. A.M.S. calculated the models of the formation and evolution of J0247-25B. A.M. conducted the investigation into the pulsation properties of J0247-25B. T.R.M. and P.F.L.M. produced the lightcurves from the Ultracam images. U.H. calculated the synthetic stellar spectra used to check our effective temperature estimates for J0247-25B. T.R.M., V.S.D., S.L. and C.C. are responsible for the operation and maintenance of Ultracam and contributed to the planning and execution of the observations. B.S. calculated the synthetic stellar spectra and performed the comparison with the observed spectra for J0247-25A. V.S. and E.B. contributed to the execution of the observations.

**Author Information** Reprints and permissions information is available at www.nature.com/reprints. Correspondence and requests for materials should be addressed to P.M. (e-mail: p.maxted@keele.ac.uk).

**Table 1 | Properties of J0247-25.**

| Parameter | J0247-25A | J0247-25B |
|---|---|---|
| Mass (solar masses) | 1.356 ± 0.007 | 0.186 ± 0.002 |
| Radius (solar radii) | 1.697 ± 0.011 | 0.368 ± 0.005 |
| Effective temperature (K) | 7,730 ± 200 | 11,380 ± 400 |
| Log luminosity (solar luminosities) | 0.97 ± 0.05 | 0.31 ± 0.06 |
| Log surface gravity (cm s$^{-2}$) | 4.111 ± 0.006 | 4.576 ± 0.011 |
| $v_{rot} \sin i$ (km s$^{-1}$) | 95 ± 5 | 30 ± 3 |
| Inclination (degrees) | 87.3 ± 0.9 | |
| Orbital period (days) | 0.6678295 ± 0.0000004 | |
| Time of mid-eclipse (TDB) | 245,4454.1066 ± 0.0002 | |
| Distance (pc) | 1,035 ± 55 | |

The masses and radii were derived from an analysis of our Ultracam photometry and UVES spectroscopy (Supplementary Notes, Supplementary Table 2, Fig. 1). The time of mid-eclipse is Barycentric Dynamical Time (TDB) given as a Julian Date. We measured the effective temperatures of the stars by fitting models to the observed flux distribution of the binary from ultraviolet to infrared wavelengths simultaneously with the constraints on the surface brightness ratio and luminosity ratio at $I_C$-band from the analysis of the eclipses (Supplementary Table 4, Supplementary Fig. 4). The projected equatorial rotation velocity $v_{rot} \sin i$ is measured from the Doppler broadening of the spectral lines (Supplementary Figs. 3 and 5).

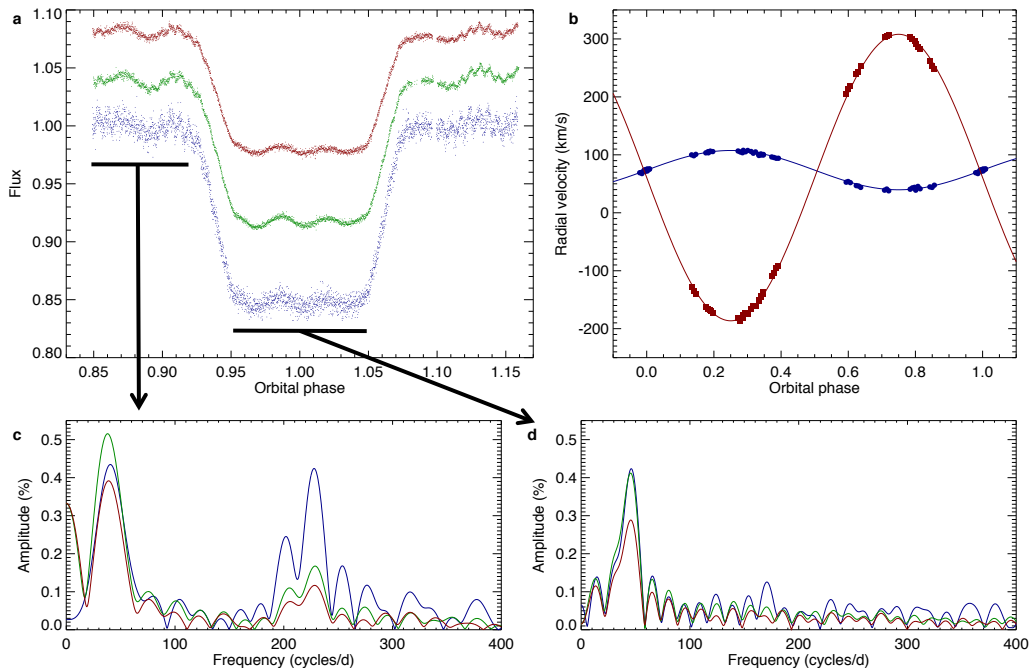

**Figure 1 | Observations of J0247-25. a,** Flux (arbitrary units) as a function of orbital phase for one primary eclipse of J0247-25 observed with Ultracam[25] on the 3.6-m NTT telescope at the European Southern Observatory (ESO). From bottom-to-top: u'-band (356 nm), g'-band (483 nm, offset by 0.04 units) and r'-band (626 nm, offset 0.08 units). Each point corresponds to an integration time of 5s. The primary eclipse is caused by the occultation of J0247-25B by the larger, cooler star J0247-25A. A secondary eclipse due to the transit of J0247-25B and a second primary eclipse were also observed (Supplementary Table 1). **b,** Radial velocity as a function of orbital phase for J0247-25A (circles) and J0247-25B (squares) measured from high resolution spectra obtained with the UVES spectrograph on the ESO 8.2-m VLT telescope. Solid lines show the predicted radial velocities for a circular orbit and our adopted values of the radial velocity semi-amplitudes, $K_A$ = 33.9 km/s and $K_B$ = 247.2 km/s. **c,** Power

spectra of the data shown in panel **a** prior to eclipse (phase range ≈ 0.85 – 0.92).  The peaks near 240 cycles/d have the highest amplitude at u'-band and the lowest amplitude at r'-band. **d,** Power spectra of the data shown in panel **a** obtained during total eclipse (phase range ≈ 0.95 – 1.05). The peaks near 220 cycles/day with amplitudes of 0.15 – 0.4% do not appear in the power spectrum during the eclipse and so must originate from J0247-25B. The pulsations near 40 cycles/day originate in J0247-25A so this is an SX Phe-type star (metal poor δ-Scuti star).

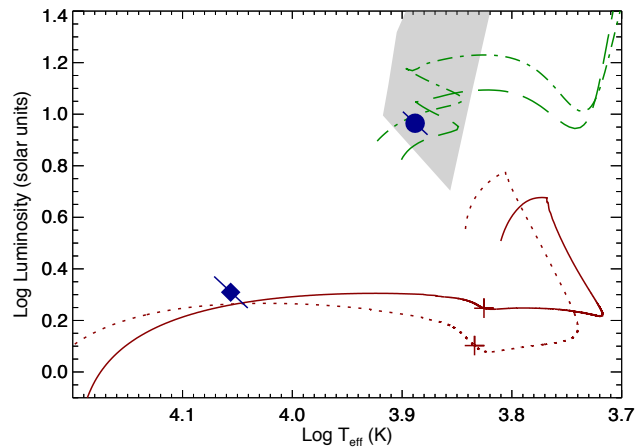

**Figure 2 | Positions of J0247-25A and J0247-25B in the Hertzsprung-Russell diagram.** J0247-25A is plotted as a filled circle, J0247-25B as a filled diamond. The error on the luminosity is correlated with the effective temperature ($T_{eff}$) so the error range (1 s.d.) is plotted as a diagonal line. Two of our models for the formation of J0247-25B are shown, one including the effects of diffusion ($Z$ = 0.004, solid line, final mass = 0.187 solar masses) and one without ($Z$ = 0.002, dotted line, final mass = 0.185 solar masses). These models are plotted from the initiation of mass transfer at log ($T_{eff}$/K) ≈ 3.8, log ($L/L_\odot$) ≈ 0.5. The cross on each track marks the end of the mass transfer phase. A model for a star with a mass of 1.35 solar masses and a composition typical for thick-disk stars ([Fe/H] = −1.0, [α/Fe] = +0.6, $Z$ = 0.005) is plotted with a dashed line[27]. A similar model for stars with [Fe/H] = 0.0 and a very high helium abundance ($Y$ = 0.4, $Z$ = 0.019) is plotted with a dot-dashed line. No match is found for any models with metal abundance outside the range $Z$ = 0.004 – 0.019. The instability strip for δ-Scuti-type pulsations with radial order $k$ = 4 is indicated by light-grey shading[28].

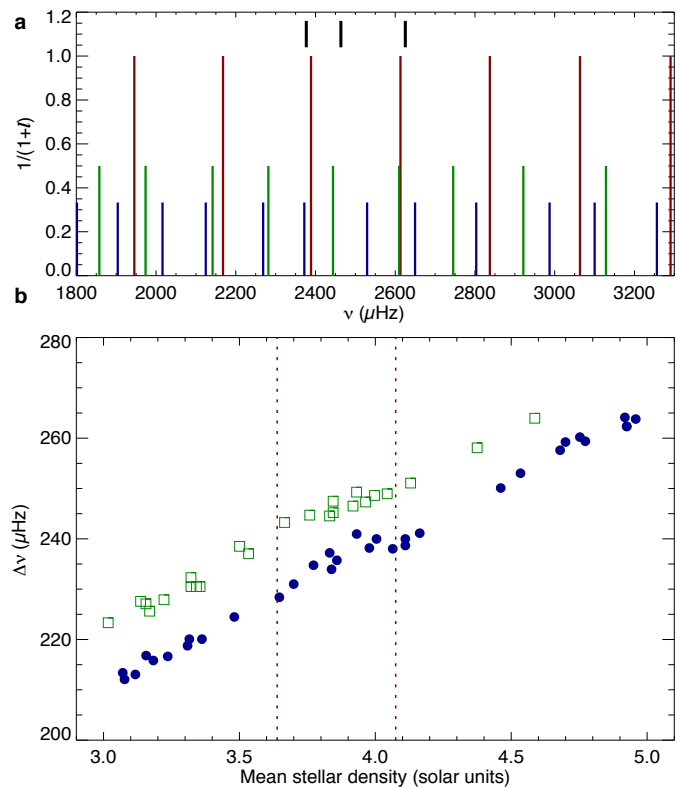

**Figure 3 | Adiabatic pulsation frequencies for models of J0247-25B. a,**
Pulsation frequencies for radial oscillation modes (ℓ = 0) and non-radial modes
with angular degree ℓ = 1 and ℓ = 2 calculated using the adiabatic approximation
for a model of a star similar to J0247-25B. Short lines at the top of the plot
indicate the observed pulsation frequencies of J0247-25B. The spacing of the
three frequencies shown cannot be explained using radial modes only. The
radial order of the modes shown is $k \approx 10$. **b,** The frequency spacing between
two adjacent radial modes, Δν, as a function of the mean stellar density for
models of J0247-25B with $Z = 0.0005 - 0.004$. The results for models with and
without diffusion are shown using filled and open symbols, respectively. The
value of Δν shown here is the mean value in the region of the theoretical power
spectrum near the observed frequencies for J0247-25B. The difference in Δν
between models with and without diffusion is a result of the sharp change in the
sound speed in the second partial ionization zone for helium due to the high helium abundance in this zone ($Y \approx 0.6$) for models without diffusion. The measured density of J0247-25B indicated using vertical dotted lines (± 1 s.d.), can be measured to high precision in this bright eclipsing binary star. As a result, models with and without diffusion can be distinguished if two of the observed pulsation frequencies are found to be radial modes.